# Prediction Modeling and Analysis for Telecom Customer Churn in Two Months

Lingling Yang[a,b], Dongyang Li[a], Yao Lu[a,b,*]

[a] School of Data and Computer Science, Sun Yat-Sen University, Guangzhou, China

[b] Guangdong Province Key Laboratory of Computational Science, 135 Xingang Xi Road, Guangzhou, China

[*] Corresponding author: luyao23@mail.sysu.edu.cn

## Abstract

A practical churn customer prediction model is critical to retain customers for telecom companies in the saturated and competitive market. Previous studies focus on predicting churn customers in current or next month, in which telecom companies don't have enough time to develop and carry out churn management strategies. In this paper, we propose a new T+2 churn customer prediction model, in which the churn customers in two months are recognized and the one-month window T+1 is reserved to carry out churn management strategies. However, the predictions for churn customers in two months are much more difficult than in current or next month because of the weaker correlation between the customer information and churn states. Two characteristics of telecom dataset, the discrimination between churn and non-churn customers is complicated and the class imbalance problem is serious, are observed. To discriminate the churn customers accurately, random forest (RF) classifier is chosen because RF solves the nonlinear separable problem with low bias and low variance and handles high feature spaces and large number of training examples. To overcome the imbalance problem, synthetic minority over-sampling with borderline or tomek link, in which the distribution of the samples remains and the number of the training examples becomes larger, is applied. Overall, a precision ratio of about 50% with a recall ratio of about 50% is achieved in the T+2 churn prediction. The proposed prediction model provides an accurate and operable churn customer prediction model for telecom companies.

# 1. Introduction

Wireless telecom licenses are granted to three mobile companies in mainland China. The annual growth rate of new mobile users decreased greatly to 4.7% in 2014 from more than 10% during the years of 2009 to 2013. The Chinese mobile telecom market is getting saturated and becomes more and more competitive, similar to that in many other countries. It is a common case now that a new customer of one telecom company is a churn customer from another one. In other words, the loss of customers usually means contributing new customers to other competitors. The life-and-death matter for a telecom company is changing from customer acquisition to customer retention [1].

Customer churn management receives a growing attention as retaining existing customers is profitable and important to telecom companies. The cost for attracting new customers is much more than the cost of retaining existing customers in business, especially in the saturated telecom market. It is reported to be five to six times more expensive[2], [3] to attract a new customer than to retain an existing customer. Moreover, the long-term customers are less sensitive to the competitive market, e.g. the long-term customers are less likely to switch to the other companies because of their promotion, and contribute more profit to the current company. Customer churn management in the framework of customer relationship management is a comprehensive strategy of recognizing churn customer, working out and implementing retention approaches. In this paper, we focus on the first aspect, predicting the customers with high possibility to leave. The efficiency of the customer churn management is proportional to the accuracy of the churn customer prediction model. It is reported that about 1.29 billion mobile users were registered in China by the end of 2014 and the number of the mobile users in China is equivalent to the sum of the mobile users in all European countries. As the number of the telecom customers is quite large, a small improvement in the churn model will contribute a great increase in the profit [4].

One contribution of our work is to propose a new churn customer prediction model which is more applicable for telecom company. Our churn customer prediction model recognizes churn customers in two months (month T+2) based on the customer information in current month (month T) and reserves the one month (month T+1) between the analysis (month T) and the outcome time window (month T+2) for telecom companies to carry out churn management strategies. To our knowledge, related work about telecom churn predictions all focus on current or next month prediction ([1], [5], [6]), such that the company will not have enough time to carry out some retention approaches. For the Chinese Unicom Guangdong Branch providing the dataset of 4 generation (4G) customers (about 2.7 million) in this paper, a suitable churn customer prediction model is quite urgent because the company suffers a month churn rate of about 7% for these 4G customers. And the month churn rate is much higher than the average month churn rate of about 2% in the telecom industry [7]. Thus for China Unicom Guangdong Branch, the T+2 churn customer prediction model is applicable and operable.

On the other hand, the difficulty of predicting churn customers in two months is much larger as the inherent churn reasons are weaker compared to the churn prediction in current or next month. Thus another contribution of our work is to achieve the best T+2 churn customer prediction performance. Two characteristics of customer churn data set, the discrimination between churn and non-churn customers is complicated and the class imbalance problem is serious, are observed. To discriminate the churn customers accurately, various machine learning algorithms, such as support vector machine (SVM) [8], decision tree [9] and logistic regression (LR) [6], have been applied to predict telecomm churn customers. We propose to use random forest (RF) because RF solves the nonlinear separable problem with low bias and low variance and handles high feature spaces as well as large number of training examples well. To overcome the class imbalance problem, synthetic minority over-sampling (SMOTE) with borderline (borderline-SMOTE) or tomek link (SMOTE+Tomeklink), is applied because the distribution of the samples will remain and the number of the training data will be larger in these two over-sampling techniques.

The paper is organized as follows. In section 2, the details of the prediction problem and data set are described. In section 3, the proposed T+2 churn customer prediction model is described. The performances of the proposed prediction models are evaluated in section 4.

And the last section is discussion part.

## 2. Materials

### 2.1. Problem Description

The T+2 churn customer prediction problem is presented in Figure 1. The customers potentially churn in two months are recognized based on user data in current month. In the figure, T, T+1 and T+2 represent the current month, the next month and two months later individually. The user data in month T (analysis window) are analyzed to predict the churn customer who will leave in two months (outcome window). The retention window (month T+1) is reserved for telecom companies to carry out churn management strategies.

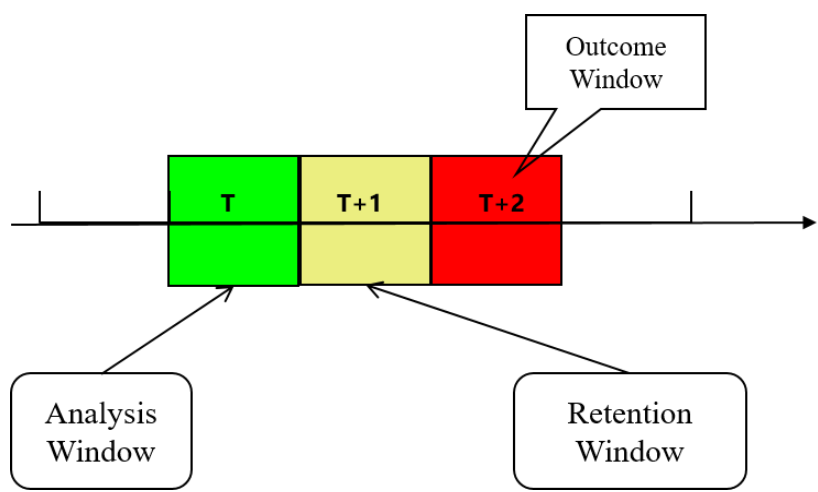

Figure 1 The T+2 Churn Prediction Problem

### 2.2. Data Set Description

The data set containing about 2.7 million 4G customers is from China Unicom Telecom Company Guangdong Branch. Here the users are the customers whose analysis windows are in month July 2015 (T = 201507), August 2015 (T = 201508), September 2015 (T = 201509) and October 2015 (T = 201510) separately. Moreover, the customers should be active all the time through months T-2, T-1 and T. For example, for the subset of T = 201507, the customers should be active through months T-2 = 201505, T-1 = 201506 and T = 201507. And for the subset of T = 201508, the customers should be active through months T-2 = 201506, T-1 = 201507 and T = 201508. Thus the ranges of the analyzed customers are slightly different from month to month. In addition, the monthly churn rate for the 4G customers is about 7%.

The variables applied in our churn prediction model are presented in Table 1. These variables are selected based on the literature study and discussion with telecom experts. The variables are classified manually into six categories, i.e. customer profiles, call details, bill details, data traffic details, month state and other information.

Table 1 Customer Variables in the T+2 Churn Prediction Model

| | | | |
|---|---|---|---|
| **Customer Profiles** | City Type | **Call Details** | Roaming Call Duration |
| | Credit | | Paid Call Duration |
| | Join Month | | Over-Product Voice Tag |
| | GAT Roaming Tag | | Domestic Long-Distance Call Duration |
| | Half Stop Flag | | GAT International Long-Distance Call Duration |
| | Provincial Roaming Tag | | Non-GAT International Long-Distance Call Duration |
| | Two-Low User Tag | | Number of Incoming Calls |
| | Three-Low User Tag | | Number of Outgoing Calls |
| | Mobile Type | **Data Traffic Details** | Paid Data Traffic |
| | TDLTE Tag | | Free Data Traffic |
| | FDDLTE Tag | | Provincial Data Traffic |
| **Bill Details** | Recharge Amount | | Domestic Data Traffic |
| | Monthly Fee | | International Data Traffic |
| | Grant Amount | | Data Traffic Used Days |
| | Arrears Amount | **Other Information** | Shutdown Days |
| | Over-Product Voice Income | | SMS Numbers |
| | Over-Product Stream Income | | Promotion Tag |
| **Month State** | Churn State at the Start of Month | | Promotion End Date |
| | Churn State at the End of Month | | |

## 3. Methodology

### 3.1. Proposed Churn Customer Prediction Model

The workflow of our proposed churn customer prediction model is illustrated in Figure 2. Monthly data in month T are first cleaned through filling in the null values and correcting the erroneous data separately. Then the cleaned monthly data in month T and the churn tags in month T+2 are put together as the training data.

It is observed that the class imbalance problem is serious as the ratio of the number of non-churn customers to the number of churn customers is generally very large. This imbalance problem [10] results in lower learning performance and is more pronounced with the application of machine learning algorithms in various fields [11]. To overcome the class imbalance problem, borderline-SMOTE or SMOTE+Tomeklink over-sampling technique, in which the distribution of the samples will remain and the number of the training data will be larger, is applied. On the other hand, the discrimination between churn and non-churn customers is complicated and the number of the data set is large (about 2.7 million 4G customers). Thus RF classifier is applied as RF solves the nonlinear separable problem with low bias and low variance and handles high feature spaces as well as large number of training examples well. The details of the borderline-SMOTE and SMOTE+Tomeklink over-sampling techniques as well as RF classifier in our proposed churn customer prediction model are described in the following paragraphs.

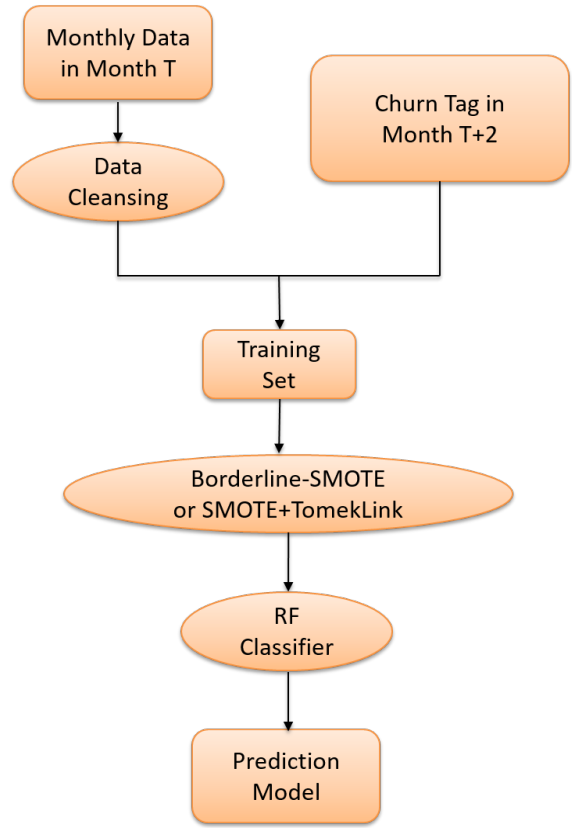

Figure 2 The Workflow of the Proposed T+2 Churn Customer Prediction Model.

*3.1.1. Borderline-SMOTE and SMOTE+Tomeklink Over-Sampling Techniques*

Typically, over-sampling techniques are applied in imbalanced data set to provide a balanced class distribution through manipulating the examples in the minority class. For conveniences, the class with more examples is called the majority class and the one with fewer examples are classed the minority class in the paper.

Synthetic minority over-sampling (SMOTE) proposed by Chawla et.al [12] is a powerful over-sampling technique which is applied to construct balanced class distributions from the unbalanced data sets in various applications [13], [14]. The SMOTE algorithm creates 'synthetic' examples based on the feature space in the minority class. The details of the algorithm are as follows. For each example $x_i \in S_{min}$ ($S_{min}$: the set of minority class examples), the $k$ minority class nearest neighbors $X_{i-neigh-min} = \{x_{ij}|x_{ij} \in S_{min}, 1 \leq j \leq k\}$ are computed based on Euclidian distance. The number of synthetic examples for each example $x_i$ is $m = (int)(M/T)$, where $M$ is the number of the whole required over-sampling minority class examples and $T$ is the number of the original minority class examples. Then $m$ neighbors $X_{i-mneigh-min}$ are randomly selected from $X_{i-kneigh-min}$ and new examples are generated along the line between $x_i$ and each $\hat{x} \in X_{i-mneigh-min}$. The new example is computed as $x_{new} = x_i + \alpha(\hat{x} - x_i)$, $\alpha \in [0,1]$ is a random number. If $M < T$, only a part of examples in the minority class will be used to create the new examples.

SMOTE generates the same number of the synthetic examples for each minority class example without considering the differences among examples, which will increase the occurrence of overlapping between classes and lead to the problem of over generalization. Borderline-SMOTE and SMOTE+TomekLink are introduced to overcome the problem.

The idea of Borderline-SMOTE [15] is to generate the synthetic examples only based on the minority class examples that are near the borderline. The details of the Borderline-SMOTE algorithm are as follows. First, the borderline minority examples are founded through checking the nearest neighbors of each minority example. For each example $x_i \in S_{min}$, the $m$ nearest neighbors $X_{m-neigh} = \{x_{ij} | x_{ij} \in S, 1 \leq j \leq m\}$ from the whole training set $S = S_{min} \cup S_{maj}$ ($S_{min}$: the set of minority class examples, $S_{maj}$: the set of majority class examples) are computed. The set of the majority class neighbors for example $x_i$ is denoted $S_{i-maj} = S_{m-neigh} \cap S_{maj}$. The example $x_i$ is considered as a dangerous example if $m/2 \leq |S_{i-maj}| < m$, which means the example $x_i$ has more majority nearest neighbors than minority nearest neighbors. While the example $x_i$ is considered as a noise if $|S_{i-maj}| = m$ and the example $x_i$ is considered as a safe example if $0 \leq |S_{i-maj}| < m/2$. Then only the dangerous minority class examples are applied to generate the synthetic examples using the SMOTE algorithm as the dangerous examples are considered to be easily misclassified.

On the other hand, SMOTE+Tomelink [16] is to solve the over generalization problem through using Tomek link as a cleaning method. Tomek link is introduced [17] to remove the overlapping between different classes. Given two examples $x_i$ and $x_j$ belong to different classes and $d(x_i, x_j)$ represents the distance between $x_i$ and $x_j$, an example pair $(x_i, x_j)$ is called a Tomek link if there is no example $x_k$, such that $d(x_i, x_k) < d(x_i, x_j)$ or $d(x_k, x_j) < d(x_i, x_j)$. The examples that form Tomek links from both classes are removed after SMOTE. Thus the synthetic minority examples which are too deeply in the majority class space will be removed.

*3.1.2. Random Forest Classifier*

Random forest is an ensemble learning method for classification and regression. Random forest randomly selects features and constructs a clustering of decision trees on the training data, which corrects the over-fitting problem of the single decision tree [18]. It is shown that the more the number of the trees trained in the random forest algorithm, the better the prediction. While the training time will increase as more decision trees are constructed. In this paper we set the number of the trees to be 100 considering both the prediction performance and the training time. Moreover, our experiments show that the prediction accuracy show minimal improvement while the training time increase greatly when the number of the trees is larger than 100. Gini index is employed as the criteria for the partition of the data to build each decision tree.

**3.2. Compared Prediction Models**

*3.2.1. LR and SVM Classifiers*

Logistic regression [6], [19], [20] and support vector machine [8], [21], [22] have been prevalently employed in telecom customer churn prediction studies.

(1) LR: Logistic regression is a regression model where the dependent variable is categorical. Logistic regression models predicts the probability that a given example belongs to the "1" class versus the probability that it belongs to the "0" class. Generally, logistic regression predicts $Y = 1$ when $p \geqslant 0.5$ and $Y = 0$ when $p < 0.5$.

(2) SVM: Support vector machine is a supervised learning algorithm for classification and regression analysis. SVM constructs a hyper-plane or sets of hyper-planes in a high dimensional space through maximizing the margin between the hyper-planes. SVM with linear kernel is applied in this paper as the feature dimensionality is high. The penalty parameter C of the error term is chosen in the range $C \in \{10^{-4}, 10^{-3}, 10^{-2}, 10^{-1}, 10^0, 10^1, 10^2, 10^3, 10^4\}$ when training prediction models in our experiments. And the results for SVM with $C = 10^2$ are shown in this paper because it achieves the highest prediction accuracy among all the values of $C$.

*3.2.2. Random/Tomek Link Under-sampling and Random Over-sampling Techniques*

(1) Random under-sampling: Random under-sampling is a non-heuristic method to

balance the class distribution of a data set through randomly removing the majority class examples.

(2) Tomek link under-sampling: If two examples form a Tomek link, either one of them is noise or both of them are near the borderline. Tomek links are used as an under-sampling technique through removing the majority class examples in this paper.

(3) Random over-sampling: Random over-sampling is also a non-heuristic method to balance the class distribution of a data set through randomly replicating the minority class examples.

*3.2.3. Cost-Sensitive Learning Methods*

Instead of balancing data distribution through different sampling strategies, cost-sensitive learning methods try to solve the imbalance learning problem by adjusting the cost of misclassification. For the imbalance learning problem, the probability of misclassifying the minority class examples is usually higher than that of majority class examples using base classifiers. Therefore, it is possible to reduce the possibility of the minority example misclassification by placing a larger penalty on misclassifying the minority class examples than that of the majority class examples.

The cost-sensitive learning method in this paper is designed based on RF. The cost-sensitive RF algorithm first assigns a weight for each class, with a higher weight for the minority class and a lower one for the majority class. Then the weight is used to calculate the Gini criterion for finding splits in the tree construction procedure. In the voting procedure, the class prediction is determined by "weighted vote", i.e. the "weighted vote" of a class is the weight for that class times the number of cases for that class. In this paper, the weight for each class is assigned automatically by the reciprocal of the ratio of minority class and majority class. For example, if the ratio of the minority class and majority class is 1:10, the weight of minority class and majority class will be 10 and 1 correspondingly.

**3.3. Performance Measures**

It's pointed out that the prediction performance for imbalance data set could not be evaluated sufficiently in terms of average accuracy [23]. The prediction model should perform well on both positive and negative examples rather than only on one class at the cost of the other class. Thus the performance of our proposed churn prediction model is measured in terms of *precision, recall, true positive rate (TPR), F-measure* and *G-mean*. These metrics are defined as:

$$precision = \frac{TP}{TP+FP}, recall = \frac{TP}{TP+FN}, TPR = \frac{TN}{TN+FP},$$

$$F-measure = \frac{2 \cdot recall \cdot precision}{recall+precision}, G-Mean = \sqrt{recall \times TNR}.$$

The meanings of $TP, FP, FN, TN$ are stated in following confusion matrix Table 2. The higher the values of all the indicators, the better the performance of the churn prediction model. Moreover churn cases are treated as positive examples and non-churn cases are treated as negative examples in this paper.

Table 2 Confusion Matrix

|  | Actual Positive | Actual Negative |
|---|---|---|
| **Predicted Positive** | TP | FP |
| **Predicted Negative** | FN | TN |

## 4. Results

The T+2 churn customer prediction results are stated in this session. The churn customers in October, November and December 2015 (month T+2) are predicted based on the user data in August, September and October 2015 (month T) separately. Moreover, each prediction model is trained over the data in the previous month. For example, the symbol ('July to August') represents that the training set is July 2015 (user data in July 2015 as variables and the user state in September 2015 as churn tag) and the testing set is August 2015 (user data in August 2015 as variables and the user state in October 2015 as churn tag). The prediction results in detail are presented in the following.

### 4.1. Prediction Results Based on Various Base Classifiers

The churn prediction results based on three base classifiers (LR, SVM and RF) are presented in Table 3. The comparisons of the average prediction results across three month predictions are shown in Figure 3. It is observed that SVM achieves the best in terms of average recall and G-mean values while the values of precision, TPR and F-measure of SVM are the lowest. On the other hand, RF achieves the best in terms of precision and TPR, thus leading to the highest F-measure values. Meanwhile, the values of recall and G-mean, especially the value of G-mean, are also relatively high for RF. Therefore, considering all the evaluation indices, RF is considered as the most suitable classifier in the first prediction model.

Table 3  Churn Prediction Results Based on Various Base Classifiers

|  |  | LR | SVM | RF |
| --- | --- | --- | --- | --- |
| **July to August** | precision | 43.10% | 12.10% | 83.83% |
|  | recall | 17.24% | 75.84% | 32.92% |
|  | TNR | 98.30% | 58.75% | 99.52% |
|  | F-measure | 0.246 | 0.209 | 0.473 |
|  | G-mean | 0.412 | 0.668 | 0.572 |
| **August to September** | precision | 67.20% | 10.00% | 78.38% |
|  | recall | 12.10% | 72.27% | 39.44% |
|  | TNR | 99.56% | 51.91% | 99.20% |
|  | F-measure | 0.205 | 0.176 | 0.525 |
|  | G-mean | 0.347 | 0.613 | 0.625 |
| **September to October** | precision | 52.89% | 18.76% | 75.59% |
|  | recall | 16.78% | 34.60% | 33.12% |
|  | TNR | 98.94% | 89.41% | 99.24% |
|  | F-measure | 0.255 | 0.243 | 0.461 |
|  | G-mean | 0.408 | 0.556 | 0.573 |

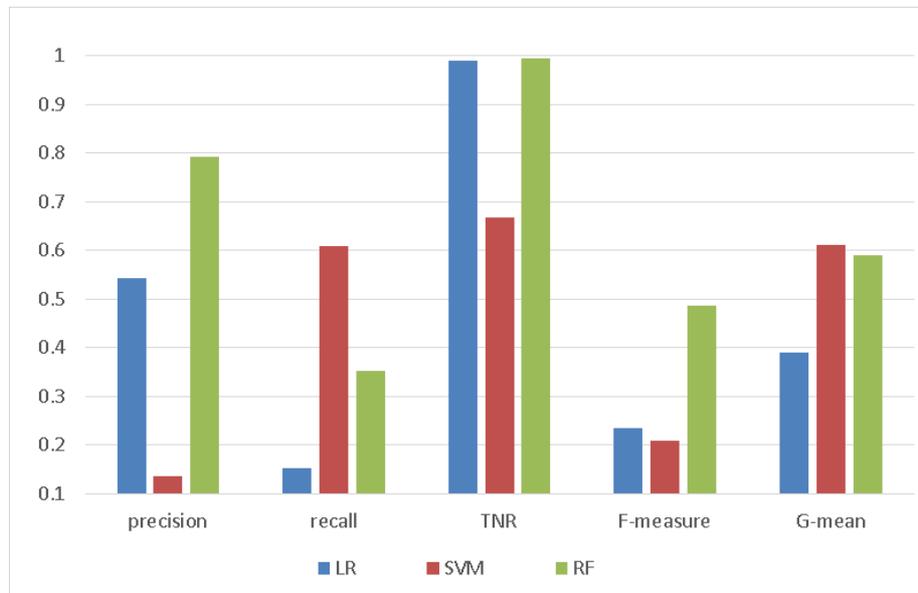

Figure 3 Average Prediction Results Based on Various Base Classifiers.

The training time of constructing the churn prediction models based on different base classifiers are presented in the Table 4. It is showed that LR takes the least time (average 80 seconds) and RF takes a little longer to construct the prediction model (average 168 seconds). While the average training time for constructing the prediction model based on SVM is 1459s (~24 minutes), which is quite time consuming.

Table 4 Training Time to Construct the Prediction Model Based on Various Base Classifiers

| Time (s) | LR | SVM | RF |
|---|---|---|---|
| July to August | 70 | 1402 | 184 |
| August to September | 74 | 1479 | 151 |
| September to October | 95 | 1495 | 169 |
| Average | 80 | 1459 | 168 |

Overall, RF outperforms LR and SVM classifiers to construct the T+2 churn prediction model considering both the prediction results and the training time. Moreover, the prediction comparisons among different sampling techniques as well as cost sensitive learning method in the following are all based on RF.

### 4.2. Prediction Results based on Sampling and Cost-sensitive Learning using RF

The churn prediction results for each individual month based on various sampling techniques and cost-sensitive learning method using random forest classifier are presented in Table 5. The average prediction results for different methods are compared in Figure 4. The base RF classifier only (no sampling) achieves an average precision of 79.27% and a TPR of 99.32%, which are the highest among all the methods, and a poor recall value, i.e. 35.16%. The average F-measure and G-mean for the base RF (no sampling) are 0.486 and 0.590 separately. The prediction recall is improved through sampling techniques or cost-sensitive learning method. It is shown that random under-sampling, cost-sensitive learning, random over-sampling, Borderline-SMOTE and SMOTE+TomekLink can improve the prediction recall greatly and achieve an average recall of 75.19%, 69.75%, 69.16%, 54.30% and 52.91% separately. On the other hand, the prediction precision decrease to 53.20% and 55.92% for Borderline-SMOTE and SMOTE+TomekLink and 28.75%, 35.39% and 36.82% for under-sampling, cost-sensitive learning, and random over-sampling individually. The prediction TPRs are always more than 90% except for under-sampling with Tomek Link. Overall, the churn prediction results are improved in terms of both F-measure and G-mean when Borderline-SMOTE and SMOTE+Tomeklink methods are applied.

Table 5 Churn Prediction Results based on Sampling and Cost-sensitive Methods using RF

| | | No Sampling | Under-Sampling | | Over-Sampling | | | Cost-Sensitive Learning |
|---|---|---|---|---|---|---|---|---|
| | | | Random Under-Sampling | Tomek Link Under-Sampling | Random Over-Sampling | Borderline-SMOTE Sampling | SMOTE + Tomek Link | |
| July to August | precision | 83.83% | 27.78% | 82.96% | 36.79% | 56.25% | 59.71% | 34.53% |
| | recall | 32.92% | 76.78% | 34.05% | 68.49% | 52.89% | 50.39% | 69.01% |
| | TNR | 99.52% | 99.48% | 85.05% | 91.19% | 97.45% | 96.92% | 90.20% |
| | F-measure | 0.473 | 0.408 | 0.483 | 0.479 | 0.545 | 0.547 | 0.460 |
| | G-mean | 0.572 | 0.808 | 0.582 | 0.790 | 0.716 | 0.701 | 0.789 |
| August to September | precision | 78.38% | 29.50% | 77.88% | 37.16% | 52.23% | 55.15% | 36.07% |
| | recall | 39.44% | 80.93% | 40.63% | 75.35% | 59.50% | 59.09% | 76.44% |
| | TNR | 99.20% | 99.15% | 85.70% | 90.58% | 96.45% | 95.98% | 89.98% |
| | F-measure | 0.525 | 0.432 | 0.534 | 0.498 | 0.556 | 0.571 | 0.490 |
| | G-mean | 0.625 | 0.833 | 0.635 | 0.826 | 0.756 | 0.755 | 0.829 |
| September to October | precision | 75.59% | 28.98% | 74.73% | 36.52% | 51.11% | 52.90% | 35.58% |
| | recall | 33.12% | 67.87% | 34.42% | 63.64% | 50.51% | 49.25% | 63.79% |
| | TNR | 99.24% | 99.18% | 88.24% | 92.18% | 96.90% | 96.58% | 91.83% |
| | F-measure | 0.461 | 0.406 | 0.471 | 0.464 | 0.508 | 0.51 | 0.457 |
| | G-mean | 0.573 | 0.774 | 0.584 | 0.766 | 0.698 | 0.691 | 0.765 |

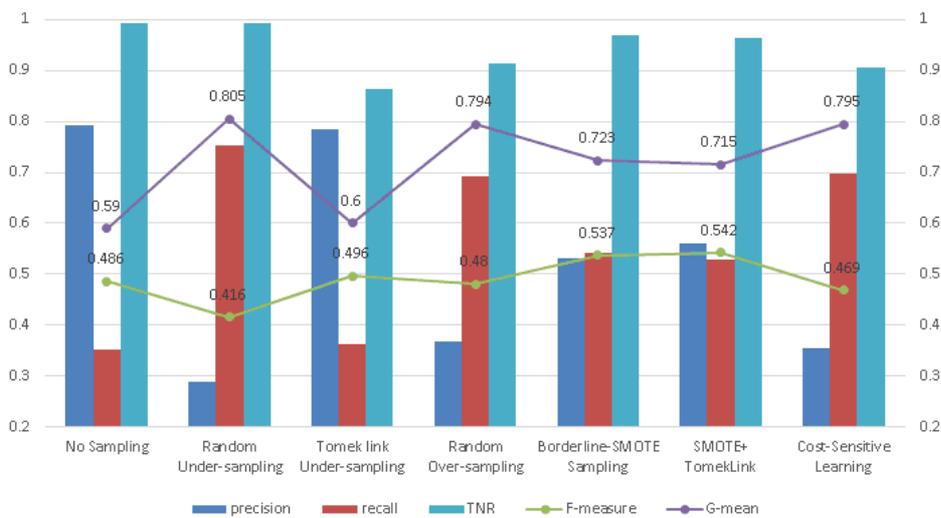

Figure 4 Average Prediction Results of Imbalanced Data Learning Methods.

The time spent on constructing the T+2 churn prediction model of various sampling techniques and cost-sensitive learning method are presented in Table 6. It is observed that the base RF with no sampling takes about 168s to construct the prediction model and the training time for the cost-sensitive learning RF is a little longer, i.e. 182s. The time for various under-sampling and over-sampling procedure varies greatly. Random under-sampling only takes 4s while SMOTE+TomekLink needs 73107s (~20h). The training procedure after random under-sampling is shorten to about 14s. While the training time for three over-sampling techniques increase to 450s, 460s and 311s separately.

Table 6 Time Spent on Constructing the Prediction Model of Different Methods

| Time (s) | | No Sampling | Under-Sampling | | Over-Sampling | | | Cost-Sensitive Learning |
|---|---|---|---|---|---|---|---|---|
| | | | Random Under-Sampling | Tomek Link Under-Sampling | Random Over-Sampling | Borderline-SMOTE Sampling | SMOTE +Tomek Link | |
| July to August | Sampling | - | 4 | 15871 | 25 | 6699 | 73107 | - |
| | Train | 184 | 12 | 185 | 513 | 488 | 290 | 170 |
| | Total | 184 | 16 | 16055 | 537 | 7187 | 73397 | 170 |
| August to September | Sampling | - | 5 | 7211 | 13 | 4658 | 59225 | - |
| | Train | 151 | 14 | 174 | 488 | 340 | 288 | 175 |
| | Total | 151 | 19 | 7385 | 500 | 4998 | 59513 | 175 |
| September to October | Sampling | - | 4 | 6410 | 7 | 3512 | 66646 | - |
| | Train | 169 | 16 | 168 | 349 | 552 | 354 | 203 |
| | Total | 169 | 19 | 6578 | 356 | 4064 | 67000 | 203 |
| Average | Sampling | - | 4 | 9830 | 15 | 4956 | 66326 | - |
| | Train | 168 | 14 | 176 | 450 | 460 | 311 | 182 |
| | Total | 168 | 18 | 10006 | 465 | 5416 | 66637 | 182 |

On the whole, sampling techniques and cost-sensitive learning method can improve the prediction results in terms of F-measure or G-mean. And Borderline-SMOTE and SMOTE+TomekLink sampling techniques perform best in terms of F-measure and G-mean, despite the over-sampling procedures take too much time. However, the sampling technique will not take extra time once the prediction model is trained.

## 5. Discussion

In this paper, we propose a new T+2 churn customer prediction model which recognizes the churn customers in two months based on customer information in current month. Compared to the current or next month prediction models, the T+2 churn customer prediction model is more applicable for telecom company as the telecom company will have an retention window (month T+1) to carry out churn management strategies. Additionally, for the Chinese Unicom Guangdong Branch providing the data set of 4 generation (4G) customers (about 2.7 million) in this paper, the T+2 churn customer prediction model is quite urgent because of the high month churn rate (about 7%). To our knowledge, current churn studies all focus on current or next month predictions and we are the first group to predict customers potentially churn in two months.

However, the predictions for churn customers in two months are much more difficult than in current or next month because of the weaker correlation between the customer information and churn states. Our proposed T+2 prediction model applies RF to classify the churn customer from the non-churn ones because RF solves the nonlinear separable problem with low bias and low variance and handles high feature spaces as well as large number of training examples well. Our results show that RF is better than LR and SVM considering both the prediction results and the training time. The reasons that RF outperforms LR and SVM are as follows. The discrimination between churn and non-churn customers is complicated, and it can't be linearly separable. Thus LR is not possible to achieve a good performance. Moreover, it is not trivial to choose appropriate hyper parameters and the kernel function of SVM for the complicated discrimination problem. On the other hand, borderline-SMOTE or SMOTE+Tomeklink is used in prediction model to overcome the class imbalance problem of the telecomm data set. For the imbalanced data set, SMOTE provides a balanced class distribution with the advantages of remaining the original distribution of the samples and enlarging the number of the training data. Moreover, the borderline or Tomeklink techniques are applied to overcome the problem of over generalization of SMOTE. The experimental results show that, compared to no sampling, the prediction model with borderline-SMOTE or SMOTE+Tomeklink recognize more churn customers with a relatively high precision and

TNR values. And borderline-SMOTE or SMOTE+Tomeklink outperforms under-sampling and cost-sensitive methods because of the larger number of training data. While that SMOTE creates 'synthetic' examples based on the feature space rather than sample space leads to the results that borderline-SMOTE or SMOTE+Tomeklink is better than random over-sampling.

What's more, the time spent on sampling and constructing the classifier are also measured in this paper. Despite the sampling time for Borderline-SMOTE and SMOTE+TomekLink is much longer than random sampling technique, it is worth mentioning that it will not take extra time once the prediction model is trained. Moreover, the training time of RF is a little bit longer than no sampling because of the larger number of training data. Additionally, it is noticed that the prediction results of under-sampling with TomekLink are almost the same as the base classifier with no sampling. And meanwhile the training time for no sampling and under-sampling with TomekLink is also similar. Based on these two results, we conclude that there are few Tomek links in our telecomm data set and there are not much difference between the original dataset and the dataset after under-sampling with TomekLink technique.

Overall, in the T+2 churn prediction, a precision ratio of about 50% with a recall ratio of about 50% are achieved based on our proposed prediction model for the telecom data set. The sampling time will not be a big problem once the prediction model is constructed. The prediction results are satisfied and is practically used to retain potential churn customers in China Unicom Telecom Company Guangdong Branch. Considering the large number of churn customers each month, our model is profitable for the telecom branch.

## Acknowledgements

The research was supported in part by the Natural Science Foundation of China under grants 11401601, by Ministry of Science and Technology of China under grant 2016YFB0200602, by Innovation Key Fund of Guangdong Province under grants 2016B030307003, 2015B010110003, 2015B020233008 and by Innovation Key Fund of Guangzhou 201604020003.

148–156.